\documentclass[prl,twocolumn,amssymb,showpacs,amsmath,nobibnotes,
superscriptaddress,floatfix, aps]{revtex4}
\usepackage{graphicx}
\usepackage{bm}

\newcommand{\be}{\begin{equation}}
\newcommand{\ee}{\end{equation}}
\newcommand{\order}{{\cal O}}
\newcommand{\fig}[1]{Fig.~\ref{#1}}
\newcommand{\alphamsb}{\alpha_{\overline{\mathrm{MS}}}}
\newcommand{\alphav}{\alpha_V}
\begin{document}
\title{High-Precision Lattice QCD Confronts Experiment}
\author{C.~T.~H.~Davies}
\author{E.~Follana}
\author{A.~Gray}
\affiliation{Department of Physics and Astronomy, University of Glasgow,
Glasgow, United Kingdom}
\author{G.~P.~Lepage}
\author{Q.~Mason}
\affiliation{Laboratory for Elementary-Particle Physics, Cornell University,
Ithaca, New York 14853}
\author{M.~Nobes}
\affiliation{Physics Department, Simon Fraser University, Vancouver, British
Columbia, Canada}
\author{J.~Shigemitsu}
\affiliation{Physics Department, The Ohio State University, Columbus, Ohio
43210}
\author{H.~D.~Trottier}
\affiliation{Physics Department, Simon Fraser University, Vancouver, British
Columbia, Canada}
\author{M.~Wingate}
\affiliation{Physics Department, The Ohio State University, Columbus, Ohio
43210}
\collaboration{HPQCD and UKQCD Collaborations}
\noaffiliation

\author{C.~Aubin}
\affiliation{Department of Physics, Washington University, St.~Louis, Missouri
63130}
\author{C.~Bernard} 
\affiliation{Department of Physics, Washington University, St.~Louis, Missouri
63130}
\author{T.~Burch} 
\affiliation{Department of Physics, University of Arizona, Tucson, Arizona
85721}
\author{C.~DeTar} 
\affiliation{Physics Department, University of Utah, Salt Lake City, Utah 84112}
\author{Steven~Gottlieb} 
\affiliation{Department of Physics, Indiana University, Bloomington, Indiana
47405}
\author{E.~B.~Gregory} 
\affiliation{Department of Physics, University of Arizona, Tucson, Arizona
85721}
\author{U.~M.~Heller} 
\affiliation{American Physical Society, One Research Road, Box 9000, Ridge, New
York 11961-9000}
\author{J.~E.~Hetrick}
\affiliation{University of the Pacific, Stockton, California 95211}
\author{J.~Osborn} 
\affiliation{Physics Department, University of Utah, Salt Lake City, Utah 84112}
\author{R.~Sugar}
\affiliation{Department of Physics, University of California, Santa Barbara,
California 93106}
\author{D.~Toussaint} 
\affiliation{Department of Physics, University of Arizona, Tucson, Arizona
85721}
\collaboration{MILC Collaboration}
\noaffiliation

\author{M.~Di Pierro}
\affiliation{School of Computer Science, Telecommunications and Information
Systems, DePaul University, Chicago, Illinois 60604}
\author{A.~El-Khadra}
\affiliation{Physics Department, University of Illinois, Urbana, Illinois
61801-3080}
\author{A.~S.~Kronfeld}
\author{P.~B.~Mackenzie}
\affiliation{Fermi National Accelerator Laboratory, Batavia, Illinois 60510}
\author{D.~Menscher}
\affiliation{Physics Department, University of Illinois, Urbana, Illinois
61801-3080}
\author{J.~Simone}
\affiliation{Fermi National Accelerator Laboratory, Batavia, Illinois 60510}
\collaboration{HPQCD and Fermilab Lattice Collaborations}
\noaffiliation

\date{31 March 2003}
\begin{abstract}
We argue that high-precision lattice QCD is now possible, for the first time,
because of a new improved staggered quark discretization. We compare a wide
variety of nonperturbative calculations in QCD with experiment, and find
agreement to within statistical and systematic errors of~3\% or less. We also
present a new determination of~$\alphamsb^{(5)}(M_Z)$; we obtain~0.121(3). We
discuss the implications of this breakthrough for phenomenology and, in
particular, for heavy-quark physics.
\end{abstract}
\maketitle
For almost thirty years precise numerical studies of nonperturbative QCD,
formulated on a space-time lattice, have been stymied by our inability to
include the effects of realistic quark vacuum polarization. In this paper we
present detailed evidence of a breakthrough that may now permit a wide variety
of  nonperturbative QCD calculations including, for example, high-precision $B$
and $D$~meson decay constants, mixing amplitudes, and semi-leptonic form
factors\,---\,all quantities of great importance in current experimental work on
heavy-quark physics.
The breakthrough comes from a new discretization for light quarks:
Symanzik-improved staggered
quarks~\cite{Bernard:1998gm,Lepage:1998id,Bernard:1998mz,Lepage:1998vj,
Orginos:1998ue,Toussaint:1998sa,Orginos:1999cr,Bernard:1999xx}.

Quark vacuum polarization is by far the most expensive ingredient in a QCD
simulation. It is particularly difficult to simulate with small quark masses,
such as $u$~and $d$~masses. Consequently, most lattice QCD (LQCD) simulations in
the past have either omitted quark vacuum polarization (``quenched QCD''), or
they have included effects for only $u$~and $d$~quarks, with masses 10--20~times
larger than the correct values. This results in uncontrolled systematic errors
that can be as large as~30\%. The Symanzik-improved staggered-quark formalism is
among the most accurate discretizations, and it is 50--1000 times more efficient
in simulations than current alternatives of comparable accuracy. Consequently
realistic simulations are possible now, with all three flavors of light quark.
An exact chiral symmetry of the formalism permits efficient simulations with
small quark masses.  The smallest $u$~and $d$~masses we use are still three
times too large, but they are now small enough that chiral perturbation theory
is a reliable tool for extrapolating to the correct masses.

In this paper we demonstrate that LQCD simulations, with this new light-quark
discretization, can deliver nonperturbative results that are accurate to within
a few percent. We do this by comparing LQCD results with experimental
measurements. In making this comparison, we restrict ourselves to quantities
that are accurately measured ($<1$\% errors), and that can be simulated
reliably with existing techniques. The latter restriction excludes unstable
hadrons and multihadron states (e.g., in nonleptonic decays); both of these are
strongly affected by the finite volume of our lattice (2.5\,fm across). Unstable
hadrons, like the $\rho$ and the $\phi$, are constantly fluctuating into
on-shell or nearly on-shell decay products that can easily propagate to the
boundaries of the lattice; similar problems afflict multihadron states.
Consequently we  focus here on hadrons that are at least~100\,MeV below decay
threshold or have negligible widths ($J/\psi$, $\Upsilon$\ldots); and we
restrict our attention to hadronic masses, and to hadronic matrix elements that
have at most one hadron in the initial and final states. These are the
``gold-plated'' calculations of LQCD\,---\,calculations that must work if LQCD
is to be trusted at all.

Unambiguous tests of LQCD are particularly important with staggered quarks.
These discretizations have the unusual property that a single quark
field~$\psi(x)$ creates four equivalent species or ``tastes'' of quark.
``Taste'' is used to distinguish this property, a lattice artifact, from true
quark flavor. A quark vacuum polarization loop in such formalisms contributes
four times what it should. To remove the duplication, the quark determinant in
the path integral is replaced by its fourth root.
This construction introduces nonlocalities that are potentially worrisome, but
much is known about the formalism that is reassuring: for example,
\begin{itemize}
\item perturbation theory, which governs the theory's short-distance behavior,
is correct to all orders;
\item phenomena, such as $\pi^0\to2\gamma$, connected with chiral anomalies are
correctly handled (because the relevant (taste-singlet) currents are only
approximately conserved);
\item the CP~violating phase transition that occurs when $m_u+m_d<0$ does
\emph{not} occur  in this formalism, but the real world is neither in this phase
nor near it;
\item the nonperturbative quark loop structure is correct up to short-distance
taste-changing interactions, which are perturbative; these interactions are
suppressed by $a^2\alpha_s$ and can be systematically
removed~\cite{Mason:2002mm}; or they can be removed after the simulation using
modified chiral perturbation theory~\cite{bernard-a2,Aubin:2002ss}. 
\end{itemize}
To press further requires nonperturbative studies. The tests we present here are
among the most stringent nonperturbative tests ever of a staggered quark
formalism (and indeed of LQCD).

The gluon configurations that we used, together with the raw simulation data for
pions and kaons, were produced by the MILC collaboration;
heavy-quark propagators came from the HPQCD collaboration.
The lattices have lattice spacings of approximately $a=1/8$\,fm and
$a=1/11$\,fm. The simulations employed an $\order(a^2)$~improved staggered-quark
discretization of the light-quark action~\cite{Lepage:1998vj}, a
``tadpole-improved'' $\order(a^2\alpha_s)$~accurate discretization of the gluon
action~\cite{Alford:1995hw}, an $\order(a^2,v^4)$~improved lattice version of
NRQCD for $b$ quarks~\cite{Davies:1995db}, and the Fermilab action for
$c$~quarks~\cite{El-Khadra:1997mp}. Several valence $u/d$~quark masses, ranging
from $m_s/2$ to~$m_s/8$, were needed for accurate extrapolations, as were sea
$u/d$~masses ranging between $m_s/2$ and~$m_s/6$. Only $u$, $d$ and $s$~quark
vacuum polarization was included; effects from $c$, $b$ and $t$~quarks are
negligible ($<1$\%) here.

To test LQCD, we first tuned its five parameters to make the simulation
reproduce experiment for five well-measured quantities. The five parameters are
the bare $u$~and $d$~quark masses, which we set equal, the bare~$s$, $c$~and
$b$~masses, and the bare QCD coupling. There are no further free parameters once
these are tuned. 

Setting $m_u=m_d$ simplifies our analysis, and has a negligible effect ($<1$\%)
on isospin-averaged quantities. We tuned the $u/d$, $s$, $c$, and $b$~masses to
reproduce measured values of $m_\pi^2$, $2m_K^2-m_\pi^2$, $m_{D_s}$, and
$m_\Upsilon$, respectively. In each case the experimental quantity is
approximately proportional to the corresponding parameter, and approximately
independent of the other parameters. 

\begin{figure}
\begin{center}
\includegraphics{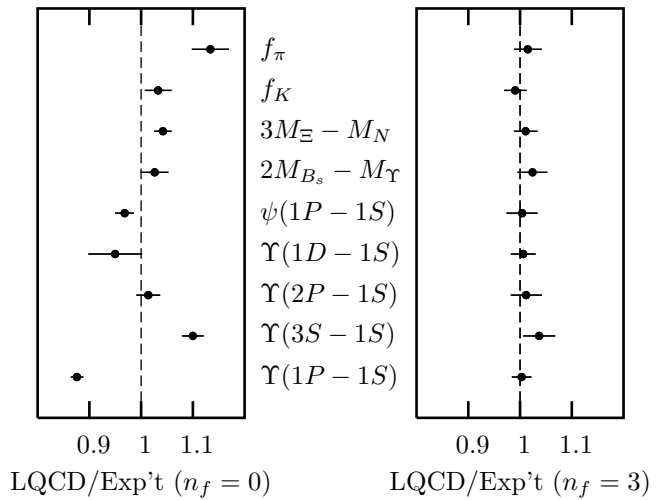}
\end{center}
\caption{LQCD results divided by experimental results for nine different
quantities, without and with quark vacuum polarization (left and right panels,
respectively). The top three results are from our $a=1/11$ and~$1/8$\,fm
simulations; all others are from $a=1/8$\,fm simulations.}
\label{ratio-fig}
\end{figure}

Rather than tune the bare coupling, one normally sets the coupling in LQCD to a
particular value, and determines the lattice spacing~$a$ in its place
(\emph{after} the simulation). We adjusted the lattice spacing to make
the~$\Upsilon$-$\Upsilon^\prime$~mass difference agree with experiment. We chose
this mass difference since it is almost independent of all quark masses,
including, in fact, the $b$~mass~\cite{Davies:1997mg}. We could equally well
have chosen, instead, any of the nine test quantities discussed below, with
similar results.

Having tuned all free parameters in the simulation, we then computed a variety
of experimentally accessible quantities (in addition to the five used for
tuning). Our results are summarized in \fig{ratio-fig} where we plot the ratio
of LQCD results to experimental results for nine quantities: $\pi$ and $K$ decay
constants, a baryon mass splitting, a $B_s$--$\Upsilon$ splitting, and mass
differences between various $J/\psi$ and $\Upsilon$ states. On the left we show
ratios from QCD simulations without quark vacuum polarization ($n_f=0$).  These
results deviate from experiment by as much as~10--15\%; the deviations can be
made as large as~20--30\% by tuning QCD's input parameters against different
physical quantities. The right panel shows results from QCD simulations that
include realistic vacuum polarization. These nine results agree with experiment
to within systematic and statistical errors of~3\% or less\,---\,with no free
parameters.
 
\begin{figure}
\includegraphics{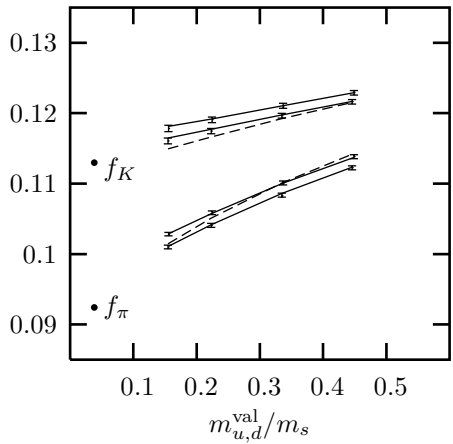}
\caption{Chiral fits to LQCD determinations of $f_\pi$ and $f_K$ (in GeV) for
different values of the valence $u/d$-quark mass at~$a=1/11$\,fm.}
\label{chiral-fig}
\end{figure}

The dominant uncertainty in the light-quark quantities in this plot (the top
four) comes from extrapolations in the sea and valence light-quark masses. We
used partially quenched chiral perturbation theory to extrapolate pion and kaon
masses, and the weak decay constants~$f_\pi$ and~$f_K$. Chiral perturbation
theory was unnecessary for correcting the $s$-quark mass; simple linear
\emph{interpolation} is adequate, and preferable since chiral perturbation
theory converges slowly for masses as large as~$m_s$.  
We also kept $u/d$~masses smaller than $m_s/2$ in our fits, so that low-order
chiral perturbation theory was sufficient. Our chiral expansions included the
full first-order contribution~\cite{Sharpe:2000bc}, and also approximate
second-order terms, which are essential given our quark masses. We corrected for
errors caused by the finite volume of our lattice (1\% errors or less), and by
the finite lattice spacing (2--3\% errors). The former corrections were
determined from chiral perturbation theory; the latter by comparing results from
the coarse and fine lattices. Residual discretization errors, due to nonanalytic
taste-violations~\cite{bernard-a2,Aubin:2002ss}, were estimated as 1.9\% for
$f_\pi$ and 1\% for~$f_K$. Perturbative matching was unnecessary for the decay
constants since they were extracted from partially conserved currents. Our final
results for $f_K$ and~$f_\pi$ agree with experiment to within systematic and
statistical uncertainties of~2.8\%. For the $n_f=0$ case we analyzed only
$a=1/8$\,fm, but extrapolated to the continuum in an approximate way based upon
our $n_f=3$~analysis.

\fig{chiral-fig}, which shows our fits for~$f_\pi$ and~$f_K$, demonstrates that
the $u/d$ masses currently accessible with improved staggered quarks are small
enough for reliable and accurate chiral extrapolations, at least for pions and
kaons. The valence and sea $s$-quark masses were 14\% too high in these
simulations; and the sea $u/d$ masses were $m_s/2.3$ and $m_s/4.5$ for the top
and bottom results in each pair. The dashed lines show the fit function with
corrected $s$ and sea-quark masses; these lines extrapolate to the final fit
results. The extrapolations are not large\,---\,only 4--9\%. Indeed the masses
are sufficiently small that simple linear extrapolations give the same results
as our fits, within few percent errors. These decay constants represent the
current state of an ongoing project; a more thorough analysis will be published
soon.

The other quantities in the ratio plot, \fig{ratio-fig}, are much less sensitive
to the valence $u/d$~mass and soft-pion effects.  Consequently, they are more
stringent tests of LQCD. The combinations $3M_\Xi-M_N$ and $2M_{B_s}-M_\Upsilon$
depend upon the valence $s$~mass, but the $s$ masses we used in our simulations
are off by only~10--20\% and easily corrected. The $b$'s rest mass cancels in
$2M_{B_s}-M_\Upsilon$, making this a particularly clean and sensitive test. The
same is true of all the $\Upsilon$ splittings, and our simulations confirm that
these are also independent ($\le1$--2\%) of the sea quark masses for our
smallest masses. The $\Upsilon(P)$~masses are averages over the known spin
states; the $\Upsilon(1D)$ is the $1^3D_2$~state recently discovered by
CLEO~\cite{Csorna:2002jg}.

It is important to appreciate that our heavy-quark results come directly from
the QCD path integral, with only bare masses and a coupling as inputs\,---\,five
numbers. Furthermore, unlike in quark models or HQET, $\Upsilon$ physics in LQCD
is inextricably linked to $B$~physics, through the $b$-quark action. Our results
strongly suggest that effective field theories, like NRQCD, are reliable and
accurate tools for analyzing heavy-quark dynamics.

Another important ingredient in high-precision LQCD is perturbation theory,
which connects lattice results to the continuum. We tested perturbation theory
by extracting values of the coupling from our simulations and comparing them
with non-LQCD results. 
We determined the renormalized coupling, 
$\alphav(6.3\,\mathrm{GeV})$, by comparing $2^\mathrm{nd}$-order 
perturbation theory for the expectation value of a $1\times1$~Wilson loop 
with (exact) values from the simulations~\cite{Davies:1997mg,Davies:2002mv}. 
Results for several sea-quark masses are shown in Table~\ref{alpha-table}; 
the masses become more realistic as one moves down the table. 

\begin{table}
\caption{\label{alpha-table} The QCD coupling $\alphav(6.3\,\mathrm{GeV})$ from
$1\times1$~Wilson loops in simulations with different $u/d$~and $s$~sea-quark
masses (in units of the physical $s$ mass), and using two different tunings for
the lattice spacing. The first error shown is statistical, and the second is
truncation error ($\order(\alphav^3)$).}
\begin{ruledtabular}
\begin{tabular}{ccccc} 
$a$ (fm) & $m_{u,d}$ & $m_s$ & $1P-1S$ & $2S-1S$ \\ \hline
1/8 & $\infty$ & $\infty$ &  0.177\,(1)( 5)  &  0.168\,(0)( 4)  \\ 
1/8 & 0.5      & $\infty$ &  0.211\,(1)( 9)  &  0.206\,(1)( 8)  \\
1/8 & 1.3      & 1.3      &  0.231\,(2)(12)  &  0.226\,(2)(11)  \\ 
1/8 & 0.5      & 1.3      &  0.234\,(2)(12)  &  0.233\,(1)(12)  \\ 
1/8 & 0.2      & 1.3      &  0.234\,(1)(12)  &  0.234\,(1)(12)  \\ 
1/11 & 0.2     & 1.1      &  0.238\,(1)(13)  &  0.236\,(1)(13)  \\ 
\end{tabular}
\end{ruledtabular}
\end{table}

The QCD coupling is sensitive to the tuning of the lattice spacing, since this
in effect tunes the bare coupling. We show results for two different tunings:
one using the $\Upsilon(1P-1S)$~splitting, and the other
using~$\Upsilon(2S-1S)$. The two tunings give couplings that are ten standard
deviations apart and 25\%~smaller than the physical coupling when the sea-quark
masses are infinite. 

With smaller, more realistic sea-quark masses, the two tunings agree to
within~1\%, and the coupling becomes mass independent. 
Our results, converted to $\overline{\mathrm{MS}}$ and evolved perturbatively to
scale~$M_Z$, imply 
\be
\alphamsb^{(5)}(M_Z) = 0.121\,(3),
\ee
which agrees well with the current world average of
$0.117\,(2)$~\cite{Hagiwara:2002fs}. Ours is the first determination from
lattice QCD simulations with realistic quark vacuum polarization, the first with
$\order(a^2)$~improved actions, and the first that is verified by a wide range
of heavy-quark \emph{and} light-quark calculations; and it is by far the most
thorough study of the light-quark mass dependence (or independence) of lattice
QCD determinations. A more detailed discussion will be presented elsewhere.

\begin{figure}[b]
        \be
        \left(
        \begin{array}{ccc}
        { \bm{V_{ud}}}   &  \bm{ V_{us}}  &   \bm{ V_{ub} }\\
        \pi\to l\nu & K\to\pi l\nu  & B\to\pi l\nu \\
        \bm{ V_{cd} }  &  \bm{ V_{cs}  } &   \bm{ V_{cb}} \\
        D\to l\nu & D_s\to l\nu & B\to D l \nu \\
        D\to \pi l\nu & D\to K l\nu  \\
        \bm{ V_{td}}  &\bm{ V_{ts}}  & \bm{ V_{tb}} \\
        \langle B_d | \overline{B}_d\rangle &
        \langle B_s | \overline{B}_s\rangle \\      
        \end{array}
        \right) \nonumber
        \ee
\caption{\label{ckm-fig}Gold-plated LQCD processes that bear on CKM matrix
elements. $\epsilon_K$ is another gold-plated quantity.}
\end{figure}

The results presented here suggest that we now have a reasonably generic and
accurate tool for solving a real-life, strongly coupled, quantum field
theory\,---\,for the first time in the history of particle physics.  Much is
required to complete the argument. Chiral extrapolations for non-strange
baryons, for example, are expected to be much larger than for pions and kaons,
as are finite-volume errors; computations with these hadrons are not yet under
control. Also a wider variety of tests is important. Heavy-quark mixing
amplitudes, and semileptonic decay form factors, for example, are essential to
the high-precision experiments at $B$~factories; our lattice techniques for
these require independent tests. The new CLEO-c program will be particularly
useful for this.

The larger challenge facing LQCD is to exploit these new techniques in the
discovery of new physics. Again, $B$~and $D$~physics offer extraordinary
opportunities for new physics from LQCD. There are, for example, gold-plated
lattice quantities for every  CKM~matrix element except $V_{tb}$
(\fig{ckm-fig}). An immediate challenge is to \emph{predict} the
$D/D_s$~leptonic and semi-leptonic decays rates to within a few percent
\emph{before} CLEO-c measures them.

This work was supported by PPARC, the National Science Foundation  and the
Department of Energy, and by computing allocations at NERSC, LANL, ORNL, SDSC,
NCSA, PSC, FNAL, and Indiana. We thank M.~Alford, T.~DeGrand, P.~Drell,
L.~Gibbons, J.~Hein, M.~Peskin, and E.~Witten for useful discussions and
comments.

\bibliography{paper.bib}

\end{document}